\documentclass[fleqn,usenatbib]{mnras}

\usepackage{newtxtext,newtxmath,bm}
\usepackage{lineno}

\usepackage[T1]{fontenc}
\usepackage{ae,aecompl}

\usepackage{graphicx}	
\usepackage{amsmath}	
\usepackage{amssymb}	
\usepackage{eso-pic}

\AddToShipoutPictureBG*{%
  \AtPageUpperLeft{%
    \hspace{0.75\paperwidth}%
    \raisebox{-4.5\baselineskip}{%
      \makebox[0pt][l]{\textnormal{Fermilab PUB-18-488-AE}}
}}}%

\title[Photo-$z$ bias from intracluster light]{Dark Energy Survey Year 1 Results:\\ The effect of intracluster light on photometric redshifts for weak gravitational lensing}

\author[Gruen et al.]{
\parbox{\textwidth}{
D.~Gruen,$^{1,2}$\thanks{E-mail: dgruen@stanford.edu; Einstein Fellow}
Y.~Zhang,$^{3}$
A.~Palmese,$^{4,3}$
B.~Yanny,$^{3}$
V.~Busti,$^{5,6}$
B.~Hoyle,$^{7,8}$
P.~Melchior,$^{9}$
C.~J.~Miller,$^{10,11}$
E.~Rozo,$^{12}$
E.~S.~Rykoff,$^{1,2}$
T.~N.~Varga,$^{7,8}$
F.~B.~Abdalla,$^{4,13}$
S.~Allam,$^{3}$
J.~Annis,$^{3}$
S.~Avila,$^{14}$
D.~Brooks,$^{4}$
D.~L.~Burke,$^{1,2}$
A.~Carnero~Rosell,$^{6,15}$
M.~Carrasco~Kind,$^{16,17}$
J.~Carretero,$^{18}$
R.~Cawthon,$^{19}$
M.~Crocce,$^{20,21}$
C.~E.~Cunha,$^{1}$
L.~N.~da Costa,$^{6,15}$
C.~Davis,$^{1}$
J.~De~Vicente,$^{22}$
S.~Desai,$^{23}$
H.~T.~Diehl,$^{3}$
J.~P.~Dietrich,$^{24,25}$
A.~Drlica-Wagner,$^{3}$
B.~Flaugher,$^{3}$
P.~Fosalba,$^{20,21}$
J.~Frieman,$^{3,19}$
J.~Garc\'ia-Bellido,$^{26}$
E.~Gaztanaga,$^{20,21}$
D.~W.~Gerdes,$^{10,11}$
R.~A.~Gruendl,$^{16,17}$
J.~Gschwend,$^{6,15}$
D.~L.~Hollowood,$^{27}$
K.~Honscheid,$^{28,29}$
D.~J.~James,$^{30}$
T.~Jeltema,$^{27}$
E.~Krause,$^{31}$
R.~Kron,$^{3,19}$
K.~Kuehn,$^{32}$
N.~Kuropatkin,$^{3}$
O.~Lahav,$^{4}$
M.~Lima,$^{5,6}$
H.~Lin,$^{3}$
M.~A.~G.~Maia,$^{6,15}$
J.~L.~Marshall,$^{33}$
F.~Menanteau,$^{16,17}$
R.~Miquel,$^{34,18}$
R.~L.~C.~Ogando,$^{6,15}$
A.~A.~Plazas,$^{35}$
A.~K.~Romer,$^{36}$
V.~Scarpine,$^{3}$
I.~Sevilla-Noarbe,$^{22}$
M.~Smith,$^{37}$
M.~Soares-Santos,$^{38}$
F.~Sobreira,$^{39,6}$
E.~Suchyta,$^{40}$
M.~E.~C.~Swanson,$^{17}$
G.~Tarle,$^{11}$
D.~Thomas,$^{14}$
V.~Vikram,$^{41}$ and
A.~R.~Walker$^{42}$
\begin{center} (DES Collaboration) \end{center}
}
\vspace{0.8cm}
\\
\\
Affiliations are listed at the end of the paper.
\vspace{1.7cm}
}

\pubyear{2018}

\begin{document}
\label{firstpage}
\pagerange{\pageref{firstpage}--\pageref{lastpage}}
\maketitle

\begin{abstract}
We study the effect of diffuse intracluster light on the critical surface mass density estimated from photometric redshifts of lensing source galaxies, and the resulting bias in a weak lensing measurement of galaxy cluster mass. Under conservative assumptions, we find the bias to be negligible for imaging surveys like the Dark Energy Survey (DES) with a recommended scale cut of $\geq200$kpc distance from cluster centers. For significantly deeper lensing source galaxy catalogs from present and future  surveys like the Large Synoptic Survey Telescope (LSST) program, more conservative scale and source magnitude cuts or a correction of the effect may be necessary to achieve per cent level lensing measurement accuracy, especially at the massive end of the cluster population.
\end{abstract}

\begin{keywords}
cosmology: observations -- galaxies: distances and redshifts -- galaxies: clusters: general -- gravitational lensing: weak
\end{keywords}



\section{Introduction}

Weak lensing mass calibration is a key to achieving the full potential of galaxy cluster cosmology \citep[for a discussion, see e.g.][]{vonderlinden2014}. Numerous lensing studies have provided cluster mass estimates over the last years \citep[e.g.][]{Gruen2014,vonderlinden2014,2015arXiv150201883H,okabesmith16,2017MNRAS.466.3103S,2017MNRAS.469.4899M,Dietrich17,2018arXiv180500039M}. The statistical power of such analyses is continuously growing with precise lensing source catalogs around large cluster samples coming from DES,\footnote{\href{https://www.darkenergysurvey.org}{https://www.darkenergysurvey.org}}, HSC,\footnote{\href{http://hsc.mtk.nao.ac.jp/ssp/}{http://hsc.mtk.nao.ac.jp/ssp/}} and KiDS\footnote{\href{http://kids.strw.leidenuniv.nl/}{http://kids.strw.leidenuniv.nl/}}, and future Euclid,\footnote{\href{https://www.euclid-ec.org/}{https://www.euclid-ec.org/}} LSST,\footnote{\href{https://www.lsst.org/}{https://www.lsst.org/}} or WFIRST\footnote{\href{https://wfirst.gsfc.nasa.gov/}{https://wfirst.gsfc.nasa.gov/}} data.
This improvement in statistics requires an equivalent push for reducing systematic uncertainties in measurement and modeling of lensing signals. State-of-the-art studies account for systematic effects such as deviations of the assumed model of the cluster matter density profile from the truth \citep[e.g.][]{2010arXiv1011.1681B}, systematics in lensing source catalogs \citep[e.g.][]{2017arXiv170801533Z}, excess contamination of the lensing source catalog with cluster member galaxies \citep[e.g.][]{2017MNRAS.469.4899M,2017arXiv170600427M}, and biases and calibration uncertainties in lensing source photometric redshifts inherent to the algorithms used for estimating them \citep[e.g.][]{2016arXiv161001160G,2017arXiv170801532H}. In recent studies, each of these effects cause uncertainty on cluster mass at the level of one to a few per cent \citep[e.g.][]{2017MNRAS.469.4899M}.

In this paper, we investigate another effect on redshift estimates of weak lensing sources -- the bias due to contamination of source photometry from diffuse intracluster light (ICL). In our ICL model, we consider light from the central galaxy and from unbound stars in the cluster potential \citep[see examples of studies or reviews in][]{1951PASP...63...61Z, 1952PASP...64..242Z, 2005ApJ...618..195G, 2005MNRAS.358..949Z, 2015IAUGA..2247903M, 2018MNRAS.474..917M, 2018AstL...44....8K} as well as the light of faint member galaxies below the survey selection threshold.
The diffuse light biases the flux and color measurements of lensing source galaxies, and causes a systematic change in photometric estimates of their redshift distributions. Among other effects, the spectral energy distribution (SED) of passive stellar populations at the cluster redshift introduces a mild cluster rest-frame D4000 break to the observed SED of the lensing source galaxy. These changes in flux and color affect the redshift assigned, especially for star-forming galaxies with weaker break features.

Careful analysis of color-magnitude space could be used to select galaxies less susceptible to these effects, and composite models for blended galaxies could in principle fully account for them. Given the complexity and algorithm dependence of source photometry and redshift estimation, we do not aim to provide a prescription for correcting ICL photo-$z$ contamination in this paper. Our goal is rather to evaluate approximately and, if possible, conservatively, what amplitude of bias we expect and identify the regimes in which it can be ignored.

In section 2, we describe our model for the surface brightness of intracluster light, using the results of \citet{zhang}. In section 3, we derive our estimate for how diffuse intracluster light of given surface brightness biases the lensing amplitude predicted from photometric redshifts, based on \citet{2016arXiv161001160G}.  Section 4 combines these two components of the model to estimate the bias in lensing excess density profiles in a typical current (DES-like) and future (LSST-like) survey, as a function of cluster redshift and separation from the cluster center. We conclude the study in section 5.

Estimates of a quantity $q$ are denoted as $\hat{q}$. All magnitudes given in the $u^{\star}g'r'i'z'$ bands are in CFHT/Megacam filters\footnote{\href{http://cfht.hawaii.edu/Instruments/Filters/megaprime.html}{http://cfht.hawaii.edu/Instruments/Filters/megaprime.html}} u.MP9301, g.MP9401, r.MP9601, i.MP9701, z.MP9801 and AB units until otherwise noted. Surface brightnesses are given in nJy arcsec$^{-2}$ units. These can be converted to counts per arcsec$^2$ at a magnitude zeropoint of 30 with a conversion factor of 3.63~nJy per count, i.e.~3.63~nJy arcsec$^{-2}$ correspond to 30 mag arcsec$^{-2}$. Cosmological distances for the scaling of lensing signal amplitudes are calculated in a flat $\Lambda$ cold dark matter cosmology with $\Omega_{m,0}=0.27$, and masses are expressed assuming a Hubble constant $H_0=70$km s$^{-1}$ Mpc$^{-1}$.

\section{Intracluster light model}
\label{sec:iclmodel}
The goal of this section is to derive a model for the surface brightness of ICL. We describe it as a function of cluster mass, cluster redshift, and projected physical distance from the cluster center.

The distribution of ICL is a debated topic in the literature. It is believed that the ICL contains a significant amount of stellar mass \citep{2013ApJ...770...57B, 2014MNRAS.437.3787C, 2018MNRAS.475..648P}, comparable to that of cluster central galaxies or the rest of the cluster galaxies. However, measurements of ICL in various samples \citep{2005MNRAS.358..949Z, 2005ApJ...618..195G, 2006AJ....131..168K, 2011MNRAS.414..602T,2012MNRAS.425.2058B, 2013ApJ...778...14G, 2014ApJ...794..137M, 2015MNRAS.448.1162D} do not necessarily find agreement on such a massive component, possibly due to methodological difference (such as differences in filter bands, or surface brightness thresholds and other criteria used to distinguish ICL from galaxies, see e.g. \citealt{2017ApJ...846..139M}, \citealt{2018MNRAS.474..917M} and the discussion in the latter), cluster-to-cluster variations \citep[e.g.,][]{2007AJ....134..466K}, cluster dynamic state \citep[e.g.,][]{2018ApJ...857...79J,2019A&A...622A.183J}, or redshift evolution and the surface brightness limits of ICL \citep[e.g.,][]{2015MNRAS.449.2353B}. By averaging the light profile of $\sim$ 300 optically selected clusters, \citet{zhang} quantified the ICL distribution at $z\sim 0.25$ for clusters more massive than $\sim 2 \times 10^{14} M_\odot$. A comparison of the stellar mass in the ICL component with the total stellar mass in DES Y1 \textsc{redMaPPer} clusters measured in \citet{palmese} shows that the ICL, together with the central galaxy, makes up $\sim 40\%$ of the total cluster stellar mass in the sample from \citet{zhang}. We make use of these measurements to model ICL.

There are three components empirically seen as diffuse light in clusters with the methodology of \citet{zhang}: pure ICL due to stars not bound to any galaxy, the light of faint cluster members below a detection/masking threshold, and scattered light of the cluster galaxies in the outskirts of the point-spread function (PSF).

We will call the first component, dominant in most regimes, \emph{pure} ICL.
Our model for pure ICL is based on the measurements presented in \citet{zhang}. In that work, sky brightness around centers of optically selected clusters is measured on co-added images. The latter are made by masking well-detected galaxies ($i<22.4$) on single-epoch DES images without background subtraction, and combining all frames of the full cluster sample while placing the cluster center at the center of the co-add image. Three effects contaminate the light measured such: background contamination due to random field galaxies, light of faint un-masked cluster member galaxies ($i>22.4$), and light of bright cluster member galaxies escaping the applied masking. These components are estimated and subtracted to yield the measurement of pure ICL.

As an additional contaminant, the PSF effect exists with every ground-based telescope at similar levels (see studies in \citealt{1969A&A.....3..455M, 1971PASP...83..199K, 1996PASP..108..699R,2007ApJ...666..663B, 2014A&A...567A..97S} and also discussions in \citealt{zhang}). It is a contaminant to the measurement in \citet{zhang}, yet greatly subdominant in the case of the DECam PSF, given that 97 per cent of light is contained within a 5'' radius of the PSF \citep[][their section 4]{zhang}, and intrinsic ICL is a much larger fraction of total cluster light. We find that the effect of PSF changes the ICL profile by less than 5 per cent in the relevant radial redshift ranges.

Our second term, the amount of light in \emph{undetected} galaxies, depends on the magnitude limit to which cluster members are detected and can be successfully deblended. We approximate this as a fixed limiting magnitude $m^{\rm lim}$.

The full function we are trying to model is thus
\begin{eqnarray}
\label{eqn:fmodel}
f_{\rm ICL}(M_{200m}, z_{d}, r, m^{\rm lim}) &=& f_{\rm pure\; ICL}(M_{200m}, z_{d}, r) \nonumber \\ &+& f_{\rm undetected}(M_{200m}, z_{d}, r, m^{\rm lim}) \; , \nonumber \\
\end{eqnarray}
with cluster mass $M_{\rm 200m}$, cluster redshift $z_{d}$, and projected physical distance $r$ from the cluster center. We describe our model for both terms in the following sections.

\subsection{Model for pure ICL}

\citet{zhang} have measured the pure ICL profile around a richness-redshift selected sample of redMaPPer clusters in DES Y1 data. In this subsection, we convert their measurement of pure ICL at these fixed parameters into a prediction for $f_{\rm pure\; ICL}(M_{200m}, z_{d}, r)$ based on the assumptions that
\begin{itemize}
\item The stellar mass density profile is self-similar, i.e.~indistinguishable between different clusters when expressed as a function of $r/r_{200m}\propto r\times M_{\rm 200m}^{-1/3}$. This is qualitatively consistent with the results of a richness-binned analysis in \citet{zhang}.
\item ICL has a fixed stellar mass density profile in physical coordinates across redshifts, which leads to a re-scaling of stellar mass per solid angle with the square of angular diameter distance. We note that there is an ongoing debate in the literature about the growth of ICL over cosmic time, which is discussed below.
\item ICL is passively evolving. As a function of redshift, it follows the corresponding luminosity evolution.
\end{itemize}

These three assumptions can be written as the three re-scaling terms on the right-hand side of the expression
\begin{eqnarray}
f_{\rm pure\; ICL}^{i'}(M_{200m}, z_{d}, r) &=& 
f_{\rm ICL}^{\rm Zhang}\left(r\times \left(\frac{M_{200m}}{M_{200m}^{\rm fid}}\right)^{-1/3}\right) \nonumber \\ 
&\times& \left(\frac{D_{A}(z_d)}{D_{A}(z_{\rm fid})}\right)^2 \nonumber \\ 
&\times& 10^{-0.4\left(m_{i',z_d}-m_{\rm fid}\right)} \; . 
\label{eqn:ftransform}
\end{eqnarray}
Here $f_{\rm ICL}^{\rm Zhang}(r)$ is the ICL surface brightness of \citet{zhang}, measured for a fiducial mass $M_{200m}^{\rm fid}=3\times10^{14} M_{\odot}$ and redshift $z_{\rm fid}=0.25$. $m_{i',z_d}-m_{\rm fid}$ is the apparent magnitude difference of a passively evolving galaxy seen at redshift $z_d$ in CFHT $i'$ band and at redshift $z_{\rm fid}$ in DES $r'$ band. For the purpose of this paper, we use a \citet{2003MNRAS.344.1000B} model with solar metallicity ($Z=0.02$), no dust, and with star formation beginning 10 Gyr before $z=0$ and subsequently declining as $e^{-t/\tau}$ with $\tau=0.1$ Gyr. The ratio of angular diameters $D_A$ corrects for the change of angular scale of the ICL profile with redshift.

\begin{figure}
	\includegraphics[width=\columnwidth]{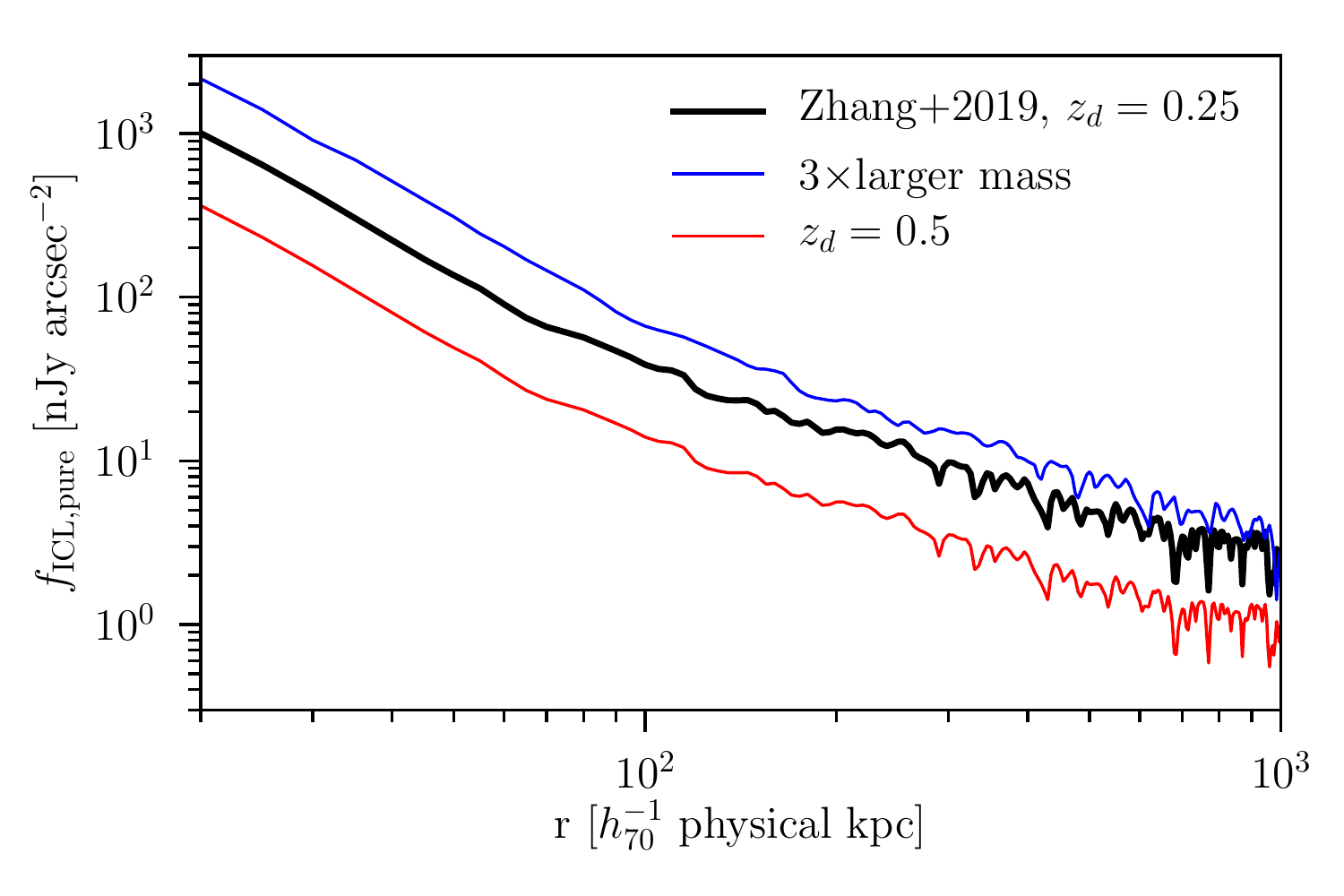}
    \caption{Pure ICL profiles (solid lines) measured in DES (black) and transformed to higher mass (blue) and redshift (red) according to \autoref{eqn:ftransform}. Dotted lines show the additional ICL due to undetected cluster members (\autoref{eqn:fundet}) in a survey that detects galaxies down to $r=22.5$.}
    \label{fig:zhangtransform}
\end{figure}

Examples of ICL profiles transformed in cluster redshift, mass and filter band are shown in \autoref{fig:zhangtransform}. In this figure and all that follows, we apply azimuthal averaging and a smoothing of $\pm40$kpc at $r>150$kpc to reduce the noise of the pure ICL measurement of \citet{zhang} at large radii.

Note that we assume ICL to not accrete or eject stars over time. It is often argued that ICL forms relatively late, assembling most of its total stellar mass during galaxy interactions after redshift 1.0 \citep{2006ApJ...648..936R,2006ApJ...652L..89M, 2007ApJ...668..826C, 2012MNRAS.425.2058B, 2013ApJ...770...57B, 2014MNRAS.437.3787C, 2016ApJ...816...98Z}. Since our model is based on ICL measurements at $z\sim 0.25$, the luminosity of ICL at higher redshift ($z>0.25$) is likely to be lower than, or at most equal to, the amount predicted from our passive evolution model. Hence, the photometric bias due to ICL at higher redshift ($z>0.25$) is likely to be less severe than that predicted in the paper. Passive evolution is a conservative assumption for the purpose of estimating photo-$z$ bias.

\begin{table}
\centering
\begin{tabular}{cccc}
$a$ & $b_r$ & $b_\lambda$&$b_z$\\
$[\mathrm{nJy}\; \mathrm{arcsec}^{-2}]$ & & & \\
\hline
$ 9.95 \pm 0.12 $ & $ 1.205 \pm 0.010 $ & $-0.831 \pm 0.037$ & $8.96 \pm 0.10$

\end{tabular}\caption{Best--fit values for eq. (\ref{eq:fmem}) for the DES $r'$ band flux from redMaPPer members. The reduced $\chi^2$ is 1.5.}\label{tab:fmem}
\end{table}

Finally, to set the ICL fluxes in other bands, we assume that at any redshift, ICL has the same color as the passive galaxy population. In reality, ICL could be somewhat bluer in color, especially at large cluster-centric distance. This is due to its lower metallicity, related to its build-up from tidal stripping of cluster members and disruption of dwarf galaxy members \citep[e.g.][]{2014ApJ...794..137M,2014A&A...565A.126P,2018MNRAS.475.3348H, 2018MNRAS.474..917M,demaio18,zhang,2019ApJ...871...24C}. We confirm that an excess in blue light of 0.1mag in $g$ band, comparable to the color effect of the expected metallicity offsets, would reduce our predicted bias, yet by less than 5 per cent, at all cluster redshifts studied here.

\subsection{ICL from undetected cluster members}

The light of undetected member galaxies is an additional contribution to diffuse light in the cluster. To add this to our full ICL model, we use the same methodology as \citet{zhang} apply for \emph{subtracting} the faint galaxy contribution towards a measurement of pure ICL (see their section 5). Namely, we assume that at any radius, the fraction of total cluster member light in faint galaxies is determined by a spatially homogeneous luminosity function. From the measured light in bright cluster members we can then predict the undetected contribution.

Formally, we write
\begin{eqnarray}
f_{\rm undetected}(M_{200m}, z_{d}, r, m^{\rm lim}) &=& f_{\rm members}(M_{200m}, z_{d}, r) \nonumber \\ &\times& S(z_d,m^{\rm lim},\infty) \; ,
\label{eqn:fundet}
\end{eqnarray}
where $S(z_d,m^{\rm lim},\infty)$ is the fraction of the integral over the cluster member luminosity function contributed by the faint end from $m^{\rm lim}$ to $\infty$.

For a \citet{1976ApJ...203..297S} luminosity function with faint-end slope $\alpha$,
\begin{equation}
\frac{\mathrm{d}N_{\rm gal}}{\mathrm{d}L}\propto\phi(L)\propto \left(\frac{L}{L^{\star}}\right)^{\alpha}\exp[L/L^{\star}]\; ,
\end{equation}
the integrated luminosity is given by
\begin{equation}
I(L_1,L_2)=\int_{L_1}^{L_2} L\;\phi(L)\,\mathrm{d}L \propto \left[\Gamma\left(\alpha+2,\frac{L_2}{L^{\star}}\right)-\Gamma\left(\alpha+2,\frac{L_1}{L^{\star}}\right)\right] \; ,
\label{eqn:lint}
\end{equation}
with the incomplete gamma function $\Gamma$. In this work, we assume $\alpha=-1$, as motivated by \citet{2014ApJ...785..104R}, thus
\begin{equation}
S(z_d,m_1,m_2)=\Gamma\left(1,10^{0.4(m^{\star}-m_1)}\right)-\Gamma\left(1,10^{0.4(m^{\star}-m_2)}\right)
\end{equation}
Note that for this luminosity function about 18\% of the total flux is contained in galaxies fainter than $0.2L_{\star}$ and more than 99\% of the total flux is contained in members brighter than $m^{\star}+5$.

The characteristic magnitude $m^{\star}$ \citep{2007ApJ...660..221K, 2014ApJ...785..104R} is a function of cluster redshift $z_d$, which we calculate from the \citet{2003MNRAS.344.1000B} model, normalized to match the SDSS DR8 \citep{2011ApJS..193...29A} redMaPPer catalog \citep{2014ApJ...785..104R} at $z=0.2$.

We approximate $f_{\rm members}$ from the light of redMaPPer cluster members $f_{\rm redMaPPer}$ in the "flux limited" DES Y1 catalog \citep{2018arXiv180500039M}. RedMaPPer estimates the probability of each galaxy along the line of sight to belong to the red cluster member population above $0.2 L_\star$ based on its position relative to the central galaxy and its color-magnitude relative to the empirically calibrated red sequence at the cluster redshift. The count of these galaxies within a cluster defines the redMaPPer richness $\lambda$. They are detected by DES over the full redshift range of the redMaPPer catalog, allowing us to empirically constrain the evolution of $f_{\rm members}$ with redshift. However, they are only the bright, red subset of the cluster galaxy population. For $f_{\rm members}$ in \autoref{eqn:fundet}, we thus use the luminosity function to re-scale $f_{\rm redMaPPer}$ by a factor $I(0.2 L^{\star},\infty)/I(0,\infty)=1.22$ from \autoref{eqn:lint}, for the missing members at $L<0.2 L_\star$. In the relevant radial range, these passive galaxies dominate the cluster member population \citep[e.g.][]{2016MNRAS.457.4360Z}, which is why we do not correct for the missing non-passive members.

From examining these light profiles, we find that the DES $r'$ band flux $f_{\rm members}$ of redMaPPer cluster members approximately follows a power law in projected radial distance, cluster richness and redshift, as:
\begin{equation}
f_{\rm members}(\lambda, z_{d}, r) = a \Big(\frac{r}{\tilde{r}} \Big)^{-b_r} \Big(\frac{\lambda}{\tilde{\lambda}}\Big)^{-b_\lambda} \Big(\frac{1+z_d}{1+\tilde{z}_d}\Big)^{-b_z} \,,\label{eq:fmem}
\end{equation}
where $\tilde{r}=240$ kpc, $\tilde{\lambda}=40$ and $\tilde{z}_d=0.5$ are the pivot values for richness and cluster redshift, and $a$ and $b_{r/\lambda/z}$ are our fit parameters for overall amplitude and power law exponents, respectively.

Eq. (\ref{eq:fmem}) is fit between 20 and 1000 kpc, and the best-fitting results from a $\chi^2$ minimization are given in Table \ref{tab:fmem} and $r$ is the comoving projected distance from the cluster center in kpc. The flux used is the SExtractor \texttt{AUTO} measurement in DES $r'$ band \citep{2017arXiv170801531D} and this is weighted for each galaxy from the redMaPPer catalog by the corresponding membership probability. The masked regions are taken into account when computing the flux per area, and the errors on the flux profiles are computed through a jackknife resampling. The bins in richness ($20<\lambda<140$) and redshift ($0.1<z<0.8$) are chosen to have a similar number of clusters in most bins. 

To convert $f_{\rm members}(\lambda, z_{d}, r)$ to $f_{\rm members}(M_{200m}, z_{d}, r)$ we apply the $\langle\ln\lambda|M_{500c}\rangle$ relation of \citet{2015MNRAS.454.2305S}. We convert between $M_{200m}$ and their $M_{500c}$ using the mass-concentration relation of \citet{2008MNRAS.390L..64D}. We note that $\langle\lambda|M\rangle\neq e^{\langle\ln\lambda|M\rangle}$ due to intrinsic scatter in $\lambda$ at fixed $M$. For the purpose of this paper and consistency with our scaling of the \citet{zhang} model for pure ICL, we set the amplitude of the scaling relation such that $\langle\lambda|M_{200m}=3\times10^{14}M_{\odot},z=0.25\rangle$=30.

The $m_{\rm lim}$ to use with \autoref{eqn:fundet} is dependent on survey and detection strategy. For the DES Y1 Gold catalog \citet[][their Figure 8]{2017arXiv170801531D}, a conservative $m_{\rm lim}$ for the purpose of estimating the contribution of cluster members to diffuse light is a DES $i'$ band magnitude of 22.5.

We model the light of undetected members at a given cluster-centric radius as homogeneously distributed, rather than concentrated at the positions of the actual galaxies. If the surface brightness of ICL at the positions of actual undetected galaxies is small enough so that the linearity of photo-$z$ bias found in \autoref{sec:dbdf} holds, the predicted mean bias does not depend on this assumption of homogeneity. For member galaxies with larger surface brightness, non-linear blending effects will likely play a role - we consider these to be an issue separate to the ICL studied in this paper.

We note that the contribution of undetected cluster members becomes important at large cluster mass, high redshift, and for a shallow survey (see dotted lines in \autoref{fig:zhangtransform}). For our DES parameters, it contributes the majority of ICL for a cluster of $M_{200m}/M_{\odot}=10^{15}$ at $z_d>0.6$. For lower mass or redshift in DES, it is a subdominant component -- contributing, in the relevant regimes, between 10 and 40 per cent of ICL. For LSST it is negligible due to the completeness down to fainter magnitudes.

\section{Lensing photo-$z$ biases from diffuse light}

The goal of this section is to derive a model for the bias in the lensing measurement of cluster surface matter density due to leakage of ICL into lensing source galaxy photometry used for estimating source redshift ($z_s$) distributions. The source redshift dependent quantity needed for lensing measurements of a matter distribution at redshift $z_d$ is the predicted amplitude of the lensing signal. This amplitude is proportional to
\begin{equation}
\beta=D_{ds}/D_{s} \; ,
\end{equation}
the ratio of angular diameter distances between lens and source $D_{ds}$ and to the source $D_{s}$, defined as the ratios of physical to angular sizes of objects at $z_s$ seen by observers at $z_d$ and 0, respectively. The true value of $\beta$ could be calculated if redshifts were known for sources and lenses. In practice, the source redshift distributions are estimated from their photometry. Any bias in photo-$z$ thus manifests as a bias in the amplitude $\hat{\beta}$ estimated from them. In this work, we therefore primarily consider biases in $\hat{\beta}$, rather than in the redshift distribution more generally.

We define this bias as
\begin{equation}
\left(\hat{\beta}/\beta\right)-1\approx F(f_{\rm ICL}, z_{d},\mathrm{source\;magnitude\;limit}) \; ,
\label{eqn:dbdf}
\end{equation}
where $f_{\rm ICL}$ is the surface brightness of intracluster light present at the position of the lensing source galaxy in question and $z_d$ is the redshift of the lens. $F$ is the model for the ICL-related bias we derive in this section. The larger the statistical power of a lensing survey, the smaller a bias can be tolerated before it significantly affects the analysis. Current (and future) surveys aim for multiplicative biases to be below the few (to one) per cent level.

In the remainder of this section we describe the basic lensing formalism, followed by our framework for estimating the impact of ICL on empirical redshift estimates in \autoref{sec:betatree}. We then develop the right-hand side of \autoref{eqn:dbdf} in \autoref{sec:dbdf}. In this, $f_{\rm ICL}$ is denoting the level of ICL surface brightness at the position of the lensing source population -- the model for $f_{\rm ICL}$ as a function of cluster mass, redshift, and distance from the cluster center was presented in \autoref{sec:iclmodel}.

The image of a lensing source (or ensemble of sources) located on some annulus around a gravitational lens at angular diameter distance $D_{d}$ from the observer is subject to tangential gravitational shear \citep[e.g.][for a review]{2001PhR...340..291B}
\begin{equation}
\gamma_t=\Sigma_{\rm crit}^{-1}\times\Delta\Sigma= \frac{4\pi G D_{d}}{c^2}\times\beta\times\Delta\Sigma \; .
\label{eq:betagamma}
\end{equation}
The excess surface density $\Delta\Sigma$ at radius $r$ is the difference of the mean mass per area \emph{inside} and \emph{on the edge} of a circle of radius $r$,
\begin{equation}
\Delta\Sigma(r)=\langle\Sigma(<r)\rangle-\Sigma(r) \; .
\end{equation}

$\hat{\beta}$ can be estimated from the photo-$z$ redshift probability density $\hat{p}(z)$ as
\begin{equation}
\hat{\beta}=\int\hat{p}(z) \frac{D_{ds}(z_d,z)}{D_s(z)}\; \mathrm{d}z \; .
\label{eqn:pz}
\end{equation}
For the mean shear signal of an ensemble of lensing source galaxies $i$, each with weight $w_i$, this can be written as
\begin{equation}
\hat{\beta}=\frac{\sum_i w_i\times\hat{\beta}_i}{\sum_i w_i} \; ,
\label{eqn:nz}
\end{equation}
where $w_i$ is a source weight and $\hat{\beta}_i$ the estimated $\beta$ of source $i$ from \autoref{eqn:pz}. For the optimal (minimum variance) estimator of mean shear or surface mass overdensity, $w_i\propto\beta_i/\sigma_{e,i}^2$, or, in practice, $\propto\hat{\beta}_i/\sigma_{e,i}^2$ where $\sigma_{e}^2$ is the shape noise variance including intrinsic and measurement noise.

In the case of an unbiased estimate $\hat{\beta}$, this connects mean tangential shear $\langle\gamma_t\rangle$ and excess surface mass density $\Delta\Sigma$ as
\begin{equation}
\langle\gamma_t\rangle=\frac{\sum_i w_i\times\gamma_{t,i}}{\sum_i w_i}=\frac{4\pi G D_{d}}{c^2}\times\hat{\beta}\times\Delta\Sigma
\end{equation}
Thus, for example, if $\hat{\beta}$ is biased low, e.g.~due to a bias in photo-$z$, the estimated $\Delta\Sigma$ is biased high, and vice versa. This is the source of bias we evaluate in the following. For an indirect impact of the bias in photometric redshifts via the estimation of cluster member contamination of the lensing source sample, see Appendix~A.

\subsection{Framework for empirical redshift estimation}
\label{sec:betatree}

Our framework for estimating the effect of ICL on photo-$z$ is a simple empirical method that gives an unbiased estimate of $p(z|\bm{m})$, where $\bm{m}$ is a vector of colors and magnitude. The accuracy of its redshift distribution estimates are limited only by selection effects or sample variance of the available reference sample with known redshifts. In this work, we use the same sample for reference and bias determination, which cancels these effects: without ICL, the redshift distribution recovery is perfect by construction. Given this, and a model for the color of and total flux from diffuse light that enters each source, we can estimate how much the $\hat{\beta}$ of \autoref{eqn:nz} will be biased. We use this simple empirical method as a proxy for any photometric redshift estimation that could be performed using similar wide-band survey data, e.g.~from DES or, with the caveat that the fainter magnitude limit is not fully covered by our CFHT-based reference catalogs, LSST.

The empirical method is a simple decision tree described in detail in \citet{2016arXiv161001160G} and publicly available at \href{https://github.com/danielgruen/betatree/}{https://github.com/danielgruen/betatree/}. Given a complete reference sample of galaxies with photometric measurements in a set of bands and with known true redshift, the decision tree provides an unbiased and close to optimal estimate of $p(z)$ based on the color-magnitude information in any subset of these bands. The method splits the color-magnitude space spanned by the subset of bands into hyper-rectangles (leaves of the decision tree), and assigns to each galaxy as its $p(z)$ the histogram of true redshifts of reference galaxies in that leaf. We make the simplifying assumption that the lensing source sample is a magnitude limited sample of galaxies, i.e.~ignore additional explicit or implicit selections on pre-seeing size, shape or profile that are commonly present in such catalogs. For the purpose of these tests, and because no sufficiently faint magnitude limited sample of galaxies with spectroscopic $z$ is available, we use the same photo-$z$ sample and (unless otherwise noted) the same settings of the tree as in \citet{2016arXiv161001160G}. The galaxies used are measured from the Canada-France-Hawaii Telescope Legacy Survey (CFHTLS) Deep fields, four fields with one sq.~deg.~area each, for which 8-band photometry from CFHTLS and the WIRCam Deep Survey (WIRDS) is available. The sample is complete to $i'\approx25$, although we use a shallower magnitude limited sample for all analyses to follow. The combination of high signal-to-noise photometry for magnitude limits relevant for lensing source samples and large volume relative to e.g.~the COSMOS field make the sample well suited for our purpose.

\bigskip
Operationally, we estimate the bias of photo-$z$ due to intracluster light with the following procedures. 

\begin{enumerate}
\item Build a decision tree from magnitude limited sample $20\leq i'\leq24$ (23.5, 24.5 as variants), optimized for a cluster redshift $z_d$, from $g'r'i'z'$ (also $u^{\star}g'r'i'z'$ as a variant) color-magnitude information. The magnitude limits are chosen to approximately match present and future samples of lensing source galaxies \citep[e.g.][]{2017arXiv170801533Z,2018PASJ...70S..25M}.
\item Estimate $\hat{\beta}$ in each leaf of the decision tree as the mean of $\beta_i$ of all reference galaxies in that leaf.
\item Determine the ICL $X-i'$ color $c_X$ as the median of the $X-i'$ color of all galaxies in the reference catalogs with $z\in[z_d-0.02,z_d+0.02]$ and a best-fit spectral energy distribution (SED) of a passive galaxy, where $X$ is one of ($u^{\star}$)$g'r'z'$. Note that this assumes that the ICL has the same SED as a red galaxy: this condition is satisfied in the clusters studied in \citet{zhang}, where the ICL colors are consistent with those from redMaPPer \citep{2014ApJ...785..104R} centrals within the inner 10 kpc, becoming bluer in the outer regions but still consistent with the red sequence galaxy population. Likewise, \citet{demaio18} found that ICL colors are consistent with red sequence galaxies over a wider redshift range ($0.29<z<0.89$) using HST imaging.
\item generate ICL-contaminated fluxes of each reference galaxy as $f_X^{\rm cont.}=f_X+\mu_A\times A\times f_{ICL,i}\times 10^{-0.4 c_X}$. In this, $A$ is defined to be the area of a circle with the post-seeing half-light radius of the galaxy. In our tests, we homogenized the data to a seeing half-light radius of 0.4'' to make this independent of the observing conditions of the CFHTLS-Deep fields. The factor $\mu_A$ accounts for the effective sensitivity of a method of measuring galaxy fluxes to diffuse light. We note that $\mu_A$ will depend strongly on the method used for extracting fluxes. By running \textsc{SExtractor} in dual-image mode with a detection image contaminated with diffuse flux, we find $\mu_{A}=2.5$ for \texttt{DETMODEL} model-fitting fluxes. \texttt{DETMODEL} fluxes are derived by fitting PSF convolved Sersic profile models to the galaxy images in a cut-out region. In our configuration, we follow the DES convention of fitting a PSF-convolved single exponential profile to the galaxies \citep{2017arXiv170801531D}. The value of $\mu_{A}=2.5$ is thus what we use in the following analysis.\footnote{In \texttt{AUTO} photometry, regardless of the explicit background subtraction settings, \textsc{SExtractor} measures and subtracts a background flux estimate locally. In this mode, it is hence insensitive to a diffuse background, i.e. $\mu_{A}^{\rm AUTO}=0$. There are other reasons, in particular the sensitivity to different point-spread function in different bands, that make \texttt{AUTO} photometry problematic for accurate multiband flux measurements in photometric surveys.}
\item Re-assign reference galaxies to leaves of the tree generated in (i), based on the contaminated color-magnitude information.
\item Estimate biased mean $\hat{\beta}$ for the contaminated case as the lensing-weighted mean (i.e.~with weight $w\propto\hat{\beta}$ of the leaf a galaxy falls into) of the respective $\hat{\beta}$ for each galaxy as determined in (ii) .
\item Estimate unbiased mean $\beta$ by weighting galaxies by their biased $\hat{\beta}$ as in (vi), but using their true reference redshifts to determine the $\beta$ to average.
\end{enumerate}

The ratio of the $\hat{\beta}$ of step (vi) and the unbiased true $\beta$ of step (vii), minus 1, is the bias we are trying to determine. Note that at $f_{\rm ICL}=0$, the two are, by construction, identical. In other words, the decision tree is an unbiased $\beta$ estimator unless the sample is affected by photometric biases or selection effects.

\subsection{Model}
\label{sec:dbdf}

In this section, we apply the scheme laid out in \autoref{sec:betatree} to derive an expression for the bias in $\Delta\Sigma$ as a function of ICL surface brightness, lens redshift, and magnitude limit of the source sample (\autoref{eqn:dbdf}).

\begin{figure}
	\includegraphics[width=\columnwidth]{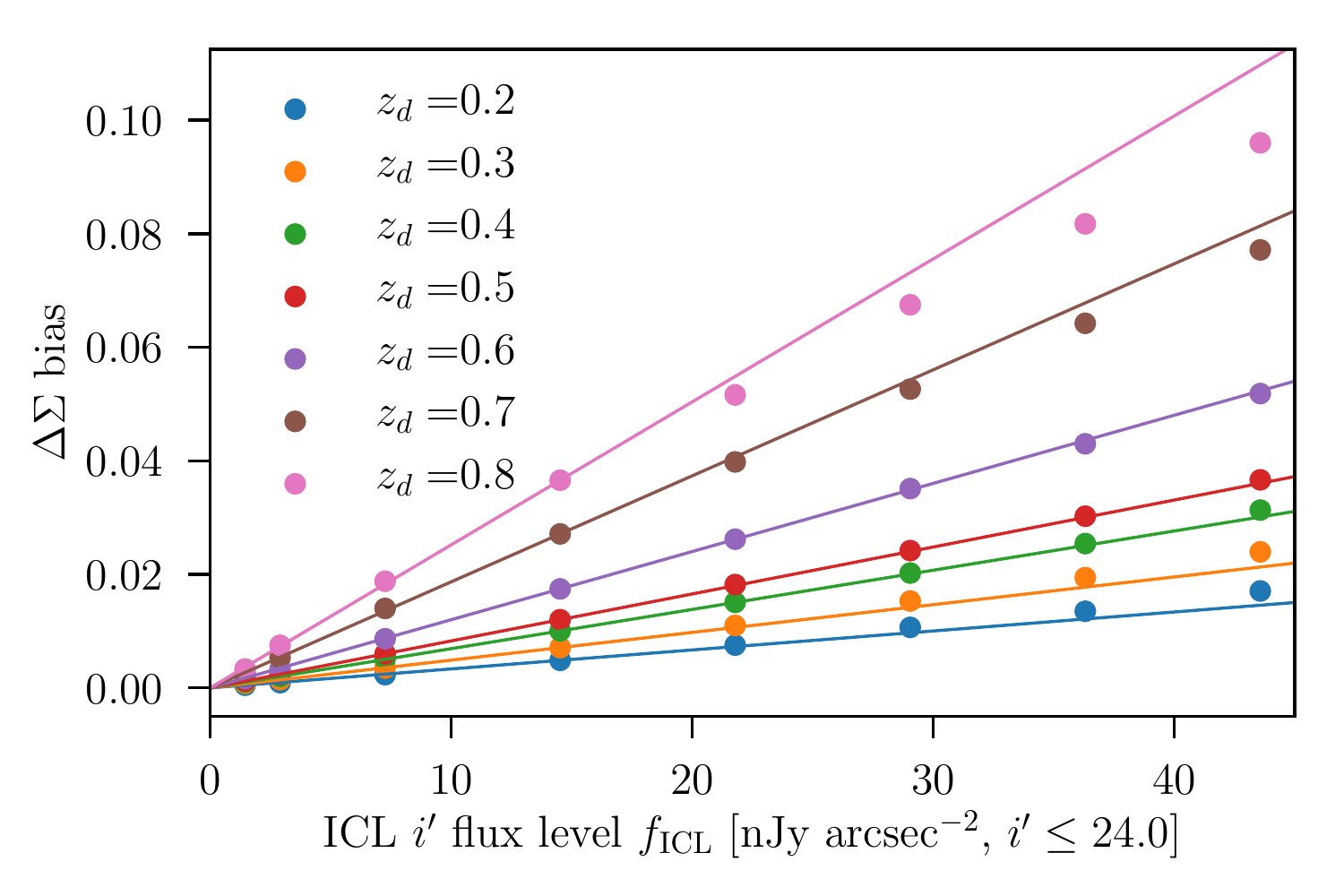}
    \caption{Bias in $\Delta\Sigma$ (defined as the negative of the bias in $\hat{\beta}$) from $g'r'i'z'$ photo-$z$ bias for a sample of source galaxies at $20\leq i'\leq24$. Differently colored lines and points show results for different lens redshifts. 3.63~nJy arcsec$^{-2}$ correspond to 30 mag arcsec$^{-2}$.}
    \label{fig:boff}
\end{figure}

Judging from the surface brightness of ICL observed in \citet{zhang}, the relevant range is $f_{\rm ICL}<40$~nJy~arcsec$^{-2}$ ($>27.4$~mag~arcsec$^{-2}$) as observed outside $\approx100$~kpc. In this range and given the sizes and magnitudes of lensing source galaxies,\footnote{Note that an $i'=24.5$ galaxy has a flux of 575~nJy, spread out over few arcsec$^2$.} ICL is a perturbation on top of the galaxies' intrinsic flux, such that we can attempt to approximate the effect of ICL on photo-$z$ as linear. We study the linearity of biases in $\hat{\beta}$ at a range of lens redshifts $z_d=0.2\ldots0.8$ in steps of 0.1 and limiting magnitudes of the source sample $m_{\rm lim}\in\{23.5,24.0,24.5\}$. \autoref{fig:boff} shows selected results for illustration that the bias is indeed well approximated as linear in $f_{\rm ICL}$ for the most relevant regimes. Only for the highest redshift clusters are non-linear effects visible at larger ICL flux levels. This is potentially related to the fact that the relevant source populations that are lensed by the cluster are located at high redshift. Their characteristic apparent magnitude is thus relatively faint and more susceptible to change due to ICL leakage. In the following, we will assume the bias on $\hat{\beta}$ due to ICL is linear in ICL flux, and use the measurement at $f_{\rm ICL}=14$~nJy~arcsec$^{-2}$ ($4$ counts per arcsec$^{-2}$ at ZP=30) to determine the slope. This choice is a trade-off: the added flux due to ICL is large enough to allow a high signal-to-noise measurement of the bias, but small enough that it does not suffer from non-linear effects or lead to problems due to sources that are below the $m<25.5$ limit of the CFHTLS-Deep catalog being boosted above the $m_{\rm lim}=23.5\ldots24.5$ magnitude limit of our source sample.

\begin{figure}
	\includegraphics[width=\columnwidth]{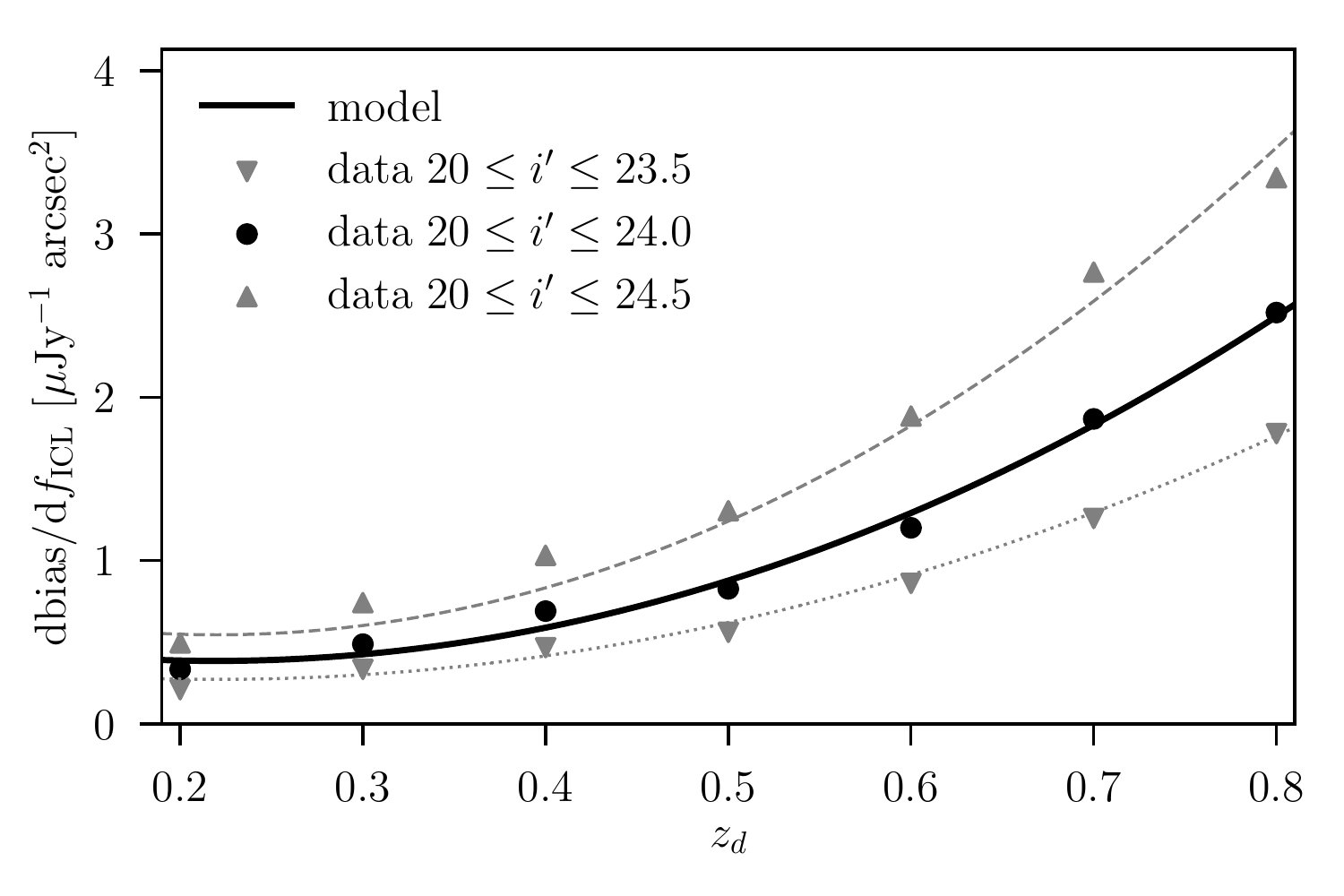}
    \caption{Slope of $\Delta\Sigma$ bias with ICL surface brightness as a function of lens redshift. Circle symbols show measurements made as in \autoref{fig:boff}, upward and downward triangles the same measurements, but for deeper and shallower source samples. Solid line shows a quadratic model fit at the fiducial magnitude limit, dashed and dotted lines are the same model re-scaled by  $2^{m_{\rm lim}-24}$, where $m_{\rm lim}$ is the limiting magnitude of the sample.}
    \label{fig:dbdf_model}
\end{figure}

For a given source magnitude limit, the slope of bias with ICL surface brightness is a function of lens redshift. By measuring the slope at a range of redshifts, we empirically find that it can be described well, within the range of $z_d=0.2\ldots0.8$, by a quadratic function of $z_d$. Measurements and quadratic model (circles and solid line) are shown in \autoref{fig:dbdf_model}.

In addition, we empirically find that a re-scaling of the model by $2^{m_{\rm lim}-24}$ describes the measurements reasonably well at magnitude limits in the range $m_{\rm lim}\in(23.5,24.5)$ (downward and upward triangles with model as dotted and dashed curve in \autoref{fig:dbdf_model}). The following is the proposed model, for $g'r'i'z'$, fitted from in $z_d\in(0.2,0.8), m_{\rm lim}\in(23.5,24.5)$,
\begin{equation}
\label{eqn:dbdfmodel}
\frac{\mathrm{d}(\hat{\beta}/\beta)}{\mathrm{d}f_{\rm ICL}}\times [\mu\mathrm{Jy}\;\mathrm{arcsec}^{-2}] \approx\left(2.5 z_d^2-1.1z_d+0.028\right)\times 2^{m_{\rm lim}-24} \; .
\end{equation}
Repeating the same analysis including $u^{\star}$ band gives a somewhat smaller amplitude of
$(1.2z_d^2-0.063 z_d +0.10)$,
to be rescaled the same way as a function of magnitude limit.

\section{Bias predictions}

Using the models described in section 2 and 3, we study the bias in $\Delta\Sigma$ profiles due to contamination of source photometry with diffuse light around clusters.

Due to the dependence on cluster member detection limit of the ICL model (section 2), and the dependence on source population of the bias per unit ICL flux (section 3), we need to define a limiting magnitude for the lensing source catalog and for the detection of cluster members in a given survey. This, in addition to the mass and redshift of a cluster sample, determines our model prediction for ICL related photo-$z$ bias via equations (\ref{eqn:dbdfmodel}) and (\ref{eqn:fmodel}).

We study two cases, and again choose conservative limits (i.e.~faint limiting magnitudes for the lensing source catalog and conservative thresholds for complete cluster member detection): (1) an ongoing $griz$ wide-area survey, similar to DES, with lensing sources measured down to $r\approx23.5$ \citep{2017arXiv170801533Z} and cluster members completely detected and deblended down to $r\approx22.5$ \citep{2017arXiv170801531D}. And (2) an ongoing or future deep wide-area $ugriz$ survey, similar to HSC or LSST, with lensing sources measured and cluster members completely detected and deblended down to $r\approx25$.

\begin{figure*}
	\includegraphics[width=\columnwidth]{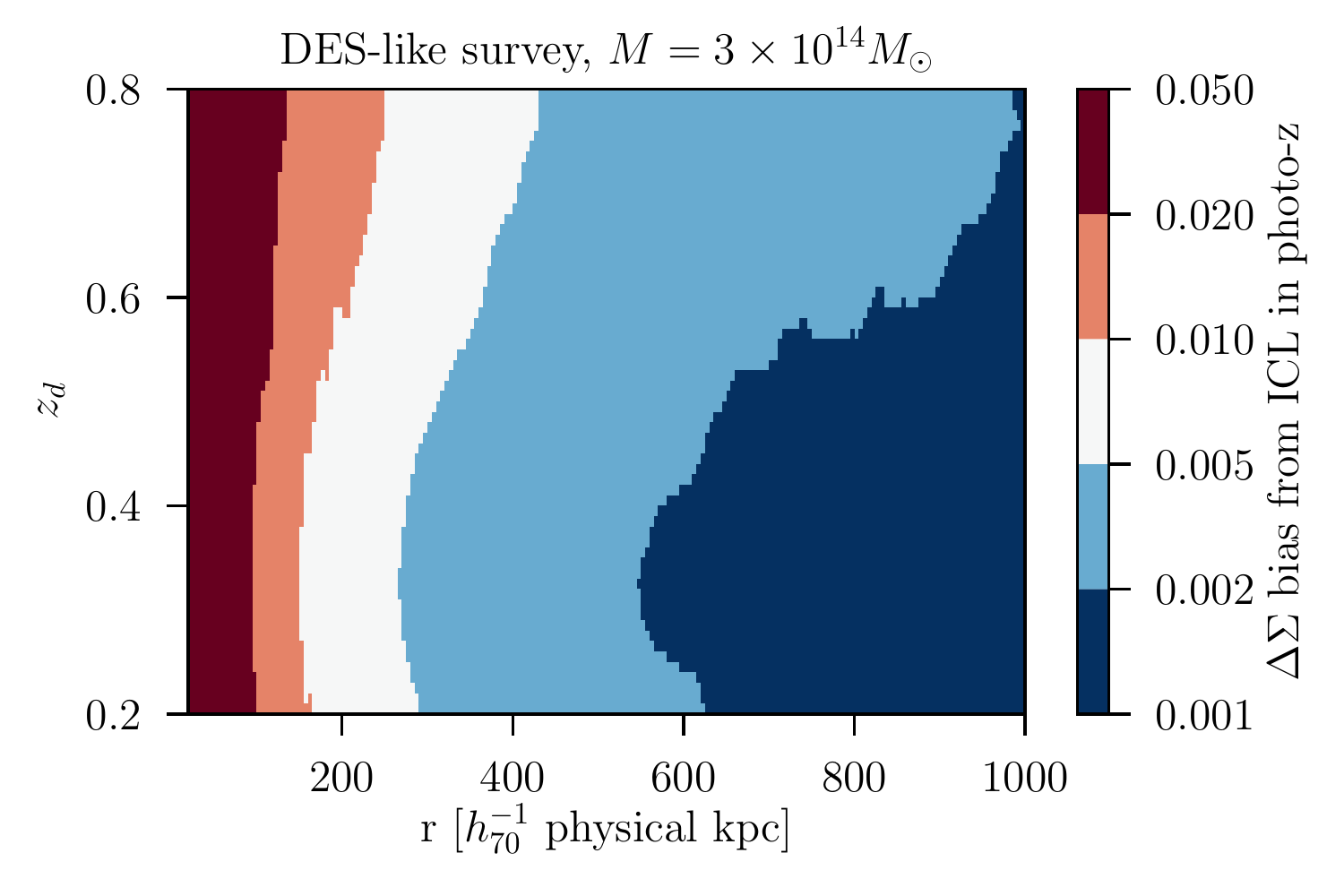}
    \includegraphics[width=\columnwidth]{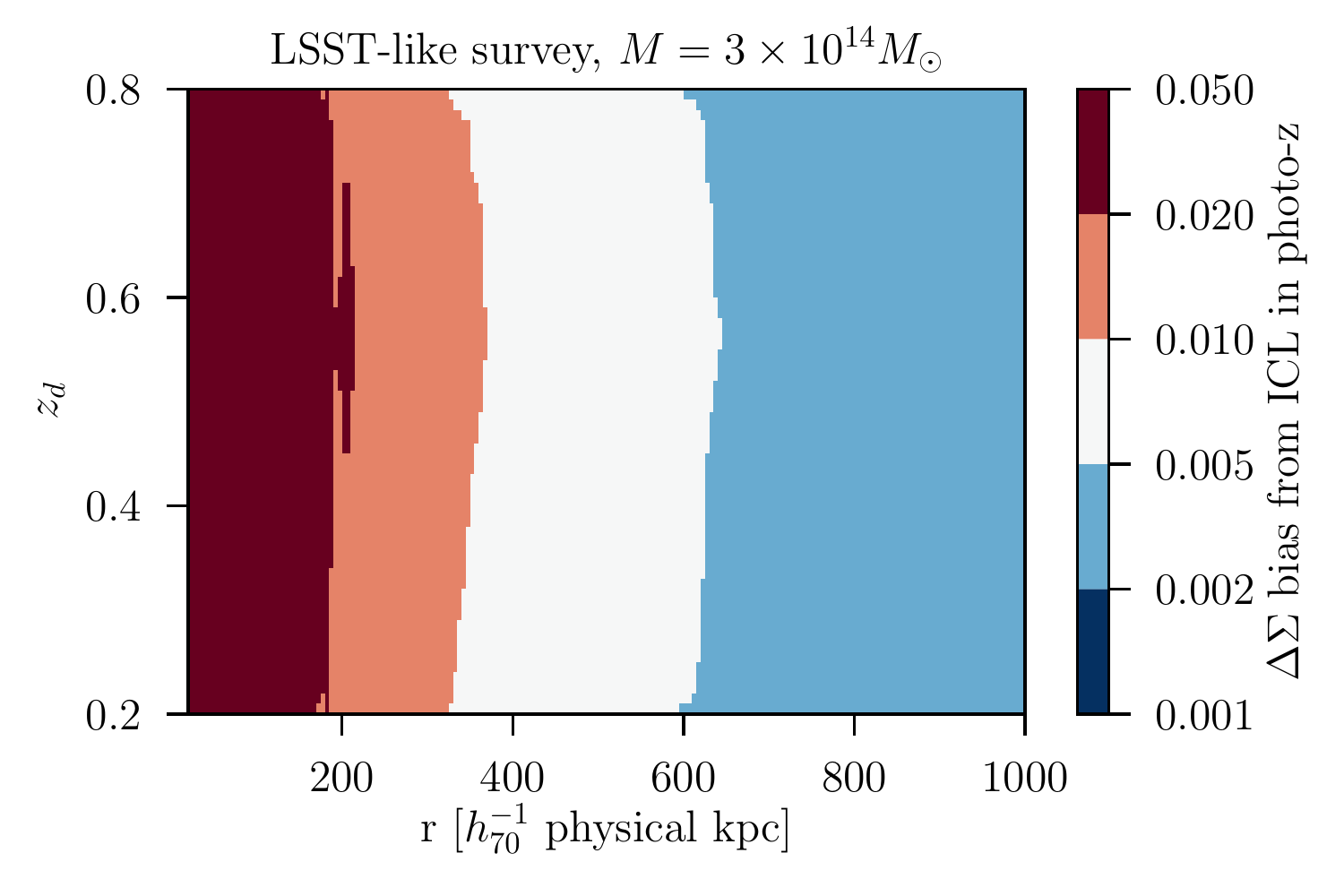}

	\includegraphics[width=\columnwidth]{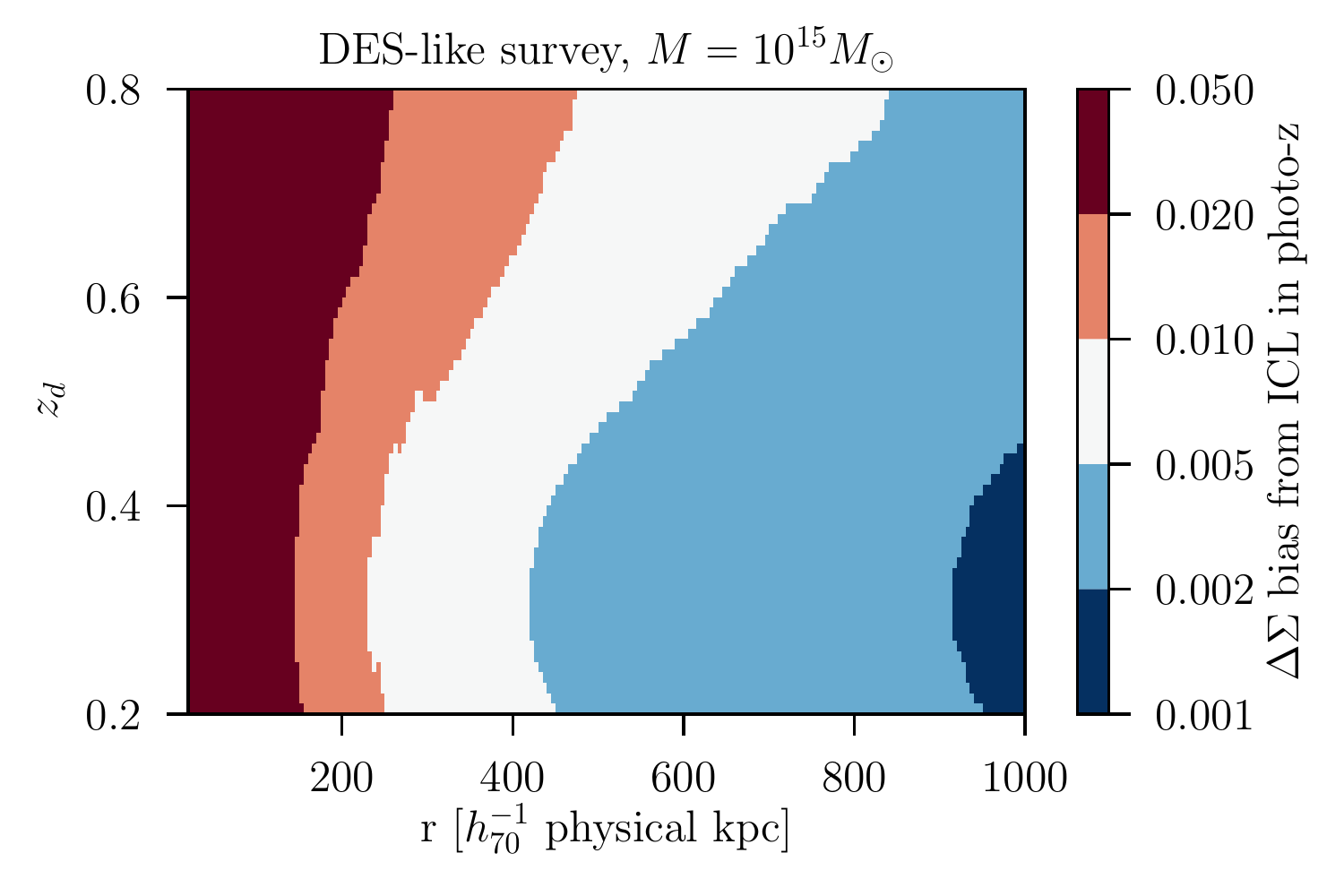}
    \includegraphics[width=\columnwidth]{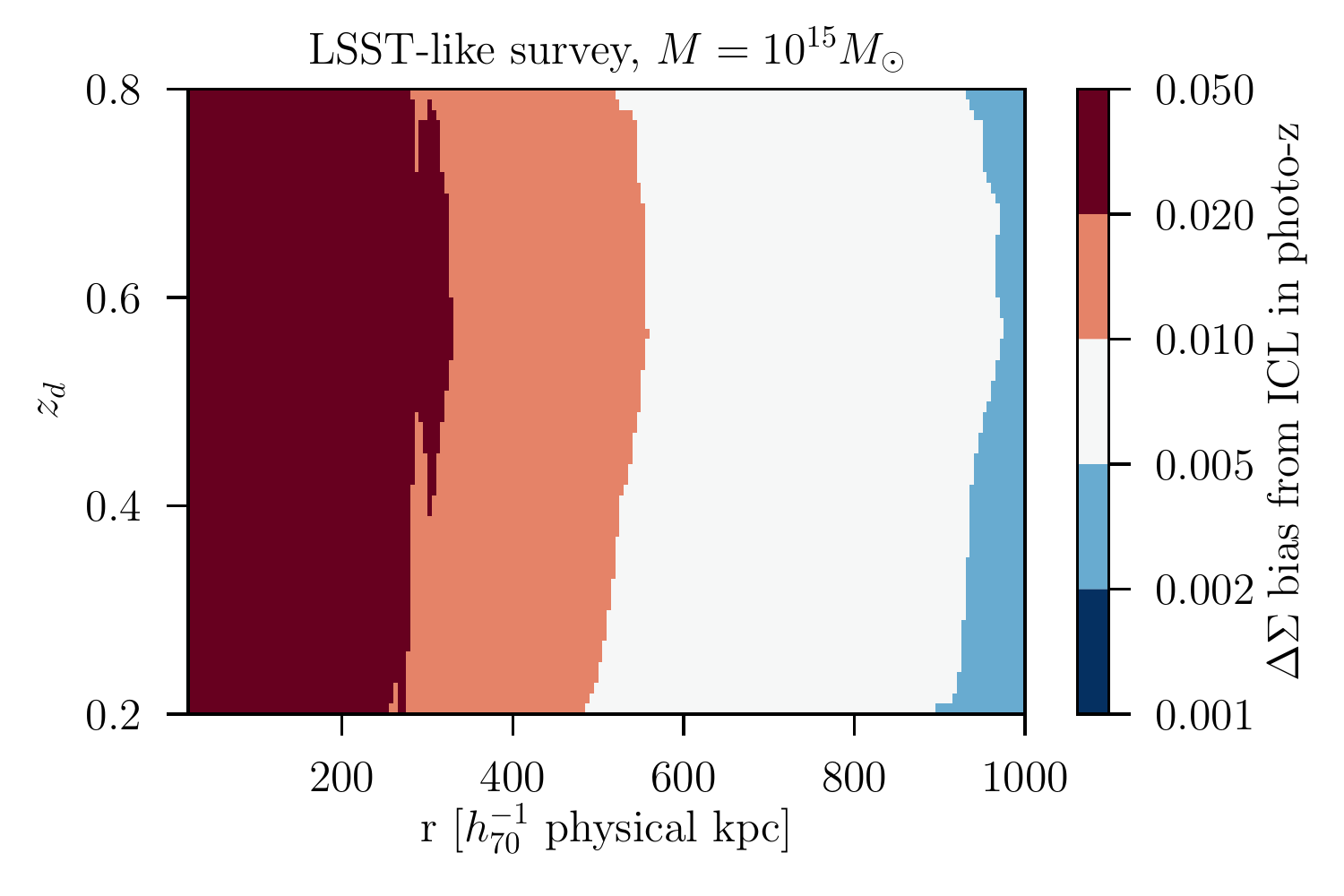}
    \caption{Predictions for the bias in $\Delta\Sigma$ profiles due to ICL-related source photo-$z$ bias for a DES-like (left-hand panels) and LSST-like (right-hand panels) survey, and a cluster of $M_{200m}/M_{\odot}=3\times10^{14}$ (top) and $10^{15}$ (bottom). The smallest scales (e.g.~$r<200 h_{70}^{-1}$ kpc in \citealt{2018arXiv180500039M}) that are most heavily affected by ICL are commonly excised from cluster lensing analyses for other reasons.}
    \label{fig:model}
\end{figure*}

Results for both cases are shown in \autoref{fig:model}, for clusters of two different masses approximately spanning the range currently used for optical cluster cosmology with \textsc{redMaPPer}. These should be compared to the statistical uncertainties of present and future surveys (currently of the order of a few per cent, optimistically of the order of one per cent) for a sense of whether the biases are relevant.

We find that for a DES-like survey, even under the conservative assumptions made above, the $\Delta\Sigma$ signal estimated outside $200$kpc radius is biased mostly below the one per cent level, and only in extreme cases above the two per cent level, even for very massive and high redshift clusters. This implies that at the scale cuts and uncertainties of present DES cluster lensing studies \citep{2018arXiv180500039M}, ICL-related photo-$z$ bias is highly subdominant compared to the 5 per cent combined statistical and systematic uncertainty.

For a significantly deeper survey like LSST, biases at the level of two per cent are possible on the small to intermediate scales of $200-300$kpc that we hope to use for cluster lensing purposes. This is driven by the larger biases incurred by the fainter sources measured in these surveys. The availability of $u$ band information in addition to $griz$ somewhat alleviates the effect. Given the conservative assumptions made in our study, it is conceivable that the actual bias is only a fraction from our model prediction. But at least for the massive end of the clusters studied with these surveys, diffuse light photo-$z$ contamination requires either more detailed investigation or more conservative cuts in radius or limiting source magnitude.

\subsection{Limitations of our model}

In the context of these predictions, we summarize the simplifications made in our model, and their likely effect on the bias in practical applications.

Simplifications, i.e.~assumptions we had to make due to limited understanding of physical or algorithmic details:
\begin{itemize}
\item \textbf{Generic photo-$z$ algorithm:} For the purpose of this test, we used a simple empirical photo-$z$ algorithm. Assuming that all photo-$z$ algorithms estimate the same relation of multiband flux and redshift, results for other algorithms would likely be similar, yet not equal. We have made simplified tests using BPZ \citep{benitez2000,2017arXiv170801532H} that indicate that this is indeed the case.
\item \textbf{Leakage of ICL into galaxy photometry:} We assumed leakage to be proportional to a circular aperture with the post-seeing half-light radius of the galaxy. This is an approximation of how a matched aperture or, equivalently, model fitting algorithm for photometric measurements might perform. While we match the leakage scale in this work to the mean observed change in \textsc{SExtractor} \texttt{DETMODEL} flux, other photometry measurement algorithms might show very different results, and galaxy morphology might affect the leakage scale in a galaxy type and redshift dependent way. Also, small scale background subtraction could greatly reduce (or even invert) the effect. It is advisable that the leakage of diffuse light into galaxy photometry is estimated from image simulations for any lensing analysis that aims at a per cent level accuracy.
\item \textbf{Linearity of $\beta$ bias as a function of ICL flux:} Our model assumes that the change in estimated lensing amplitude $\hat{\beta}$ is linear in the ICL surface brightness. While this is appropriate for the relevant range of mean ICL surface brightness, inhomogeneity (i.e.~due to undetected yet localized cluster members) could affect the photo-$z$ more or less than predicted here. At the level of deblending possible with present and future lensing surveys, we expect this effect to be subdominant.
\item \textbf{Pure red cluster member population:} We have assumed that the cluster galaxy population only contains passive galaxies, similar in color to the ones identified by the \textsc{redMaPPer} algorithm. In practice, clusters contain star forming galaxies, especially at lower mass and higher redshift. The light of the undetected members among them is likely to have a similar, but not quite equal, effect on photo-$z$ bias as the light of red members. On the radial scales considered here, star-forming members are, however, not a majority of the population. In addition, the light of undetected cluster members is a subdominant component relative to pure ICL, hence we do not expect this assumption to significantly change our conclusions.
\item \textbf{Self-similar scaling of pure ICL:} We have assumed that pure ICL scales self-similarly with cluster mass, i.e.~its surface brightness is fixed at a given projected $r/r_{500}$. While this is consistent with simple comparisons made in \citet{zhang}, a more detailed study could reveal deviations.
\end{itemize}

Conservative assumptions, i.e.~ways in which we likely overestimate the effect of ICL in practice:
\begin{itemize}
\item \textbf{Passive SED of ICL:} We assume ICL to share the color of passive galaxies at the cluster redshift. A population of younger stars in the ICL would likely reduce its effect on photo-$z$ bias due to its similarity in color to lensing sources at higher redshift. We find that predicted bias is reduced, yet by less than 5 per cent, if ICL should be brighter in $g$ band by 0.1mag, which is approximately the level expected from reduced metallicity.
\item \textbf{Lack of ICL growth}: We fix the ICL surface density to a measurement at low redshift and predict the expected bias at higher redshift without accounting for any growth of ICL from early to late times. If ICL is assembled over time we thus overestimate biases for higher redshift clusters.
\item \textbf{Conservative deblending limits:} For DES, we have assumed cluster members to be deblended and thus not affecting source photo-$z$ down to a magnitude limit of $r=22.5$. At this level, DES Y1 is highly complete -- a significant fraction of cluster members below this limit are likely deblended successfully and, unlike assumed, do not in fact contribute to diffuse ICL. As a result, we likely overestimate the associated photo-$z$ bias in DES, in particular at large cluster mass and redshift.
\item \textbf{Magnitude limited source sample:} We used a simple magnitude cut to define our source sample. Realistic lensing source samples have additional selection criteria. A choice of limiting magnitude at the faint end of the population that is used in a given analysis allows for a conservative prediction of potential biases. For DES Y1/Y3 data, this was possible to do in this work.
\end{itemize}

Limitations, i.e.~regimes in which our model is not reliable:
\begin{itemize}
\item \textbf{Faint limit of source sample:} For LSST data, sufficiently faint reference samples of galaxies with known redshift and flux measurements do not exist to extend the modeling beyond $i'\approx 25$. Assuming that fainter lensing samples are used, the bias derived here is an underestimate of the bias encountered by such analyses.
\item \textbf{Blending with cluster members:} We only attempt to model diffuse ICL leaking into source photometry at a subdominant level. For the effect of blending between similarly bright cluster member and lensing source galaxies, the model developed here is not applicable. Besides, the success of correctly treating these cases will likely strongly depend on the choice of deblending algorithm.
\end{itemize}

\section{Conclusions}

We have developed a model for the bias in weak lensing estimates of cluster surface mass overdensity due to the contamination of lensed galaxy photometry from diffuse intracluster light. The latter systematically changes the flux, color, and thus photometric redshift estimate of the faint galaxies used as lensing sources.

Our model for diffuse light in clusters is simplistic yet conservative for the purpose of this exercise: a pure component of ICL due to un-bound stars in the cluster potential, measured at low redshift \citep{zhang} and re-scaled in mass and redshift by assuming self-similarity and passive evolution; and a component due to stars in undetected, faint cluster members, extrapolated from detected galaxies by means of the luminosity function. The effect of this surface brightness on photo-$z$ is estimated from an idealized empirical photo-$z$ estimation scheme \citep{2016arXiv161001160G}.

We find that for a DES-like cluster lensing experiment, i.e.~with cluster masses up to $M_{200m}=10^{15}M_{\odot}$, detection and deblending of cluster members brighter than $i'=22.5$, and a source sample no fainter than $i'=23.5$, ICL-related photo-$z$ bias does not significantly affect weak lensing mass reconstruction. Outside a cluster-centric radius of $200$kpc, which is commonly excluded in lensing studies for other reasons, biases are typically below 1 per cent for an $M_{200m}~3\times10^{14}M_{\odot}$ cluster, and below 2 per cent at $M_{200m}~10^{15}M_{\odot}$, even under the conservative assumptions we make. The effect of ICL on measured galaxy shapes may well be larger than that, and should be tested with dedicated image simulations.

Deeper source catalogs will be somewhat more susceptible to ICL-related photo-$z$ biases because the flux and color of faint source galaxies can be changed more strongly by ICL contamination. For massive clusters, lensing source catalogs down to $i'=25$ show one per cent biases at approximately twice the radius as the above DES-like survey. Even fainter sources will likely show even stronger effects, although this is difficult to quantify at present due to the lack of reliable color-magnitude-redshift information for such samples. An explicit treatment of measured fluxes as a composite of intracluster and lensing source galaxy light in photo-$z$ estimation could in principle remedy this effect. With moderately conservative scale and magnitude cuts, however, ICL bias of photo-$z$ will be a non-issue even in the next generation of surveys -- and with a less conservative examination of the effect, these could likely be moderately relaxed from the recommendations given in this work.

\section*{Acknowledgements}

Support for DG was provided by NASA through the Einstein Fellowship
Program, grant PF5-160138 and by Chandra Award Number GO8-19101A, issued by the Chandra X-ray Observatory Center. This work was supported in part by the U.S. Department of Energy under 
contract number DE-AC02-76SF00515.

The authors thank Fabrice Brimioulle for providing the CFHTLS Deep photometry and photo-$z$ catalogs used in this work. This work has been promoted by fruitful discussions during the workshop series ``Becoming a One-Percenter''.

Funding for the DES Projects has been provided by the U.S. Department of Energy, the U.S. National Science Foundation, the Ministry of Science and Education of Spain,
the Science and Technology Facilities Council of the United Kingdom, the Higher Education Funding Council for England, the National Center for Supercomputing
Applications at the University of Illinois at Urbana-Champaign, the Kavli Institute of Cosmological Physics at the University of Chicago,
the Center for Cosmology and Astro-Particle Physics at the Ohio State University,
the Mitchell Institute for Fundamental Physics and Astronomy at Texas A\&M University, Financiadora de Estudos e Projetos,
Funda{\c c}{\~a}o Carlos Chagas Filho de Amparo {\`a} Pesquisa do Estado do Rio de Janeiro, Conselho Nacional de Desenvolvimento Cient{\'i}fico e Tecnol{\'o}gico and
the Minist{\'e}rio da Ci{\^e}ncia, Tecnologia e Inova{\c c}{\~a}o, the Deutsche Forschungsgemeinschaft and the Collaborating Institutions in the Dark Energy Survey.

The Collaborating Institutions are Argonne National Laboratory, the University of California at Santa Cruz, the University of Cambridge, Centro de Investigaciones Energ{\'e}ticas,
Medioambientales y Tecnol{\'o}gicas-Madrid, the University of Chicago, University College London, the DES-Brazil Consortium, the University of Edinburgh,
the Eidgen{\"o}ssische Technische Hochschule (ETH) Z{\"u}rich,
Fermi National Accelerator Laboratory, the University of Illinois at Urbana-Champaign, the Institut de Ci{\`e}ncies de l'Espai (IEEC/CSIC),
the Institut de F{\'i}sica d'Altes Energies, Lawrence Berkeley National Laboratory, the Ludwig-Maximilians Universit{\"a}t M{\"u}nchen and the associated Excellence Cluster Universe,
the University of Michigan, the National Optical Astronomy Observatory, the University of Nottingham, The Ohio State University, the University of Pennsylvania, the University of Portsmouth,
SLAC National Accelerator Laboratory, Stanford University, the University of Sussex, Texas A\&M University, and the OzDES Membership Consortium.

Based in part on observations at Cerro Tololo Inter-American Observatory, National Optical Astronomy Observatory, which is operated by the Association of
Universities for Research in Astronomy (AURA) under a cooperative agreement with the National Science Foundation.

The DES data management system is supported by the National Science Foundation under Grant Numbers AST-1138766 and AST-1536171.
The DES participants from Spanish institutions are partially supported by MINECO under grants AYA2015-71825, ESP2015-66861, FPA2015-68048, SEV-2016-0588, SEV-2016-0597, and MDM-2015-0509,
some of which include ERDF funds from the European Union. IFAE is partially funded by the CERCA program of the Generalitat de Catalunya.
Research leading to these results has received funding from the European Research
Council under the European Union's Seventh Framework Program (FP7/2007-2013) including ERC grant agreements 240672, 291329, and 306478.
We  acknowledge support from the Australian Research Council Centre of Excellence for All-sky Astrophysics (CAASTRO), through project number CE110001020, and the Brazilian Instituto Nacional de Ci\^encia
e Tecnologia (INCT) e-Universe (CNPq grant 465376/2014-2).

This manuscript has been authored by Fermi Research Alliance, LLC under Contract No. DE-AC02-07CH11359 with the U.S. Department of Energy, Office of Science, Office of High Energy Physics. The United States Government retains and the publisher, by accepting the article for publication, acknowledges that the United States Government retains a non-exclusive, paid-up, irrevocable, world-wide license to publish or reproduce the published form of this manuscript, or allow others to do so, for United States Government purposes.

Based in part on observations obtained with MegaPrime/MegaCam, a joint project of CFHT and CEA/IRFU, at the Canada-France-Hawaii Telescope (CFHT) which is operated by the National Research Council (NRC) of Canada, the Institut National des Science de l'Univers of the Centre National de la Recherche Scientifique (CNRS) of France, and the University of Hawaii. This work is based in part on data products produced at Terapix available at the Canadian Astronomy Data Centre as part of the Canada-France-Hawaii Telescope Legacy Survey, a collaborative project of NRC and CNRS.

This paper has gone through internal review by the DES collaboration.




\bibliographystyle{mnras}
\bibliography{literature} 




\appendix

\section{Effect of ICL on boost factor estimates}
\label{sec:boost}
Leakage of cluster members into the lensing source sample, i.e.~the erroneous use of cluster members as putative background galaxies, is a well-known cause for systematic error in cluster lensing. Because cluster members are not gravitationally lensed regardless of their estimated redshift, this reduces the amplitude of the measured shear signal relative to a model prediction. Many analyses, especially those suffering from relatively poor photometric information that does not allow a pure selection of lensing sources at $z>z_{\rm cl}$ without great losses in sample size, have used a radially dependent \emph{boost factor} correction \citep{2004AJ....127.2544S}. That is, they divided the measured signal (or multiplied the model prediction) by a factor equal to the fraction of lensing weight actually due to non-member galaxies \citep[e.g.][]{2017MNRAS.469.4899M,2018arXiv180500039M}.

The quantity needed for this correction is the fraction of lensing weight due to cluster members $f_{\rm cl}$ in each radial bin. This has often been estimated from the clustering of lensing sources with the lens positions. The blending of sources with large, bright cluster member galaxies is a known contaminant that is, however, difficult to quantify and correct without full re-processing of the survey with artificially injected faint galaxy images.

A different way of finding $f_{\rm cl}$ is based on the decomposition of the estimated, lensing weighted $p_{\rm est}(z)$ into a component measured in non-cluster fields $p_{\rm field}(z)$ and a component with different shape due to contaminating cluster members $p_{\rm member}(z)$, as
\begin{equation}
p_{\rm est}(z)=(1-f_{\rm cl})\times p_{\rm field}(z)+f_{\rm cl}\times p_{\rm member}(z) \; .
\label{eqn:boost}
\end{equation}
This method, developed in a series of papers \citep{Gruen2014,2017MNRAS.469.4899M,varga2018} and applied in several other works \citep{2017arXiv170600427M,Dietrich17,Chang17,Stern18} has the advantage that it is at first order insensitive to blending. It is, however, potentially susceptible to photo-$z$ biases and source redshift dependent selection effects in the vicinity of the cluster (see also the note in \citealt{2017arXiv170600427M}, their section 6.2).

We test the effect of ICL leakage into photometry on boost factors estimated with \autoref{eqn:boost}. Specifically, we use the scheme implemented in \citet{2017MNRAS.469.4899M} and \citet{2018arXiv180500039M} and validated in \citet{varga2018} to check the methodology of these studies in the presence of ICL. Here, $p_{\rm member}$ is assumed to be a Gaussian distribution. Its mean and width are varied, alongside $f_{\rm cl}$, to find the best-fitting boost factor in a least-squared metric between the left and right side of \autoref{eqn:boost}.

We simulate the presence of a member population of a cluster at redshift $z_{d}$ by mixing the redshift distribution of a magnitude limited sample of $i'<23.5$ with a Gaussian of mean $z_{d}+0.1$ and width $\sigma_z=0.1$.

For true contaminations $f_{\rm cl}=0.1,0.2,0.4$ and lens redshifts between $z_d\in[0.2,0.6]$, common for the settings in \citep{2017MNRAS.469.4899M,2018arXiv180500039M}, the maximum bias introduced by ICL in our model at $f_{\rm ICL}=15$nJ arcsec$^{-2}$ is $\Delta f_{\rm cl}=0.008$, or
\begin{equation}
\frac{\mathrm{d}f_{\rm cl}}{\mathrm{d}f_{\rm ICL}}\lesssim 0.0005 \; .
\end{equation}
This is to be interpreted as a multiplicative bias on $\Delta\Sigma$ and significantly smaller than the effect on $\beta$ shown in \autoref{fig:boff}. Where the latter is negligible, ICL does therefore not significantly impact boost factors estimated from $p(z)$ decomposition.

\section*{Affiliations}
$^{1}$ Kavli Institute for Particle Astrophysics \& Cosmology, P. O. Box 2450, Stanford University, Stanford, CA 94305, USA\\
$^{2}$ SLAC National Accelerator Laboratory, Menlo Park, CA 94025, USA\\
$^{3}$ Fermi National Accelerator Laboratory, P. O. Box 500, Batavia, IL 60510, USA\\
$^{4}$ Department of Physics \& Astronomy, University College London, Gower Street, London, WC1E 6BT, UK\\
$^{5}$ Departamento de F\'isica Matem\'atica, Instituto de F\'isica, Universidade de S\~ao Paulo, CP 66318, S\~ao Paulo, SP, 05314-970, Brazil\\
$^{6}$ Laborat\'orio Interinstitucional de e-Astronomia - LIneA, Rua Gal. Jos\'e Cristino 77, Rio de Janeiro, RJ - 20921-400, Brazil\\
$^{7}$ Max Planck Institute for Extraterrestrial Physics, Giessenbachstrasse, 85748 Garching, Germany\\
$^{8}$ Universit\"ats-Sternwarte, Fakult\"at f\"ur Physik, Ludwig-Maximilians Universit\"at M\"unchen, Scheinerstr. 1, 81679 M\"unchen, Germany\\
$^{9}$ Department of Astrophysical Sciences, Princeton University, Peyton Hall, Princeton, NJ 08544, USA\\
$^{10}$ Department of Astronomy, University of Michigan, Ann Arbor, MI 48109, USA\\
$^{11}$ Department of Physics, University of Michigan, Ann Arbor, MI 48109, USA\\
$^{12}$ Department of Physics, University of Arizona, Tucson, AZ 85721, USA\\
$^{13}$ Department of Physics and Electronics, Rhodes University, PO Box 94, Grahamstown, 6140, South Africa\\
$^{14}$ Institute of Cosmology \& Gravitation, University of Portsmouth, Portsmouth, PO1 3FX, UK\\
$^{15}$ Observat\'orio Nacional, Rua Gal. Jos\'e Cristino 77, Rio de Janeiro, RJ - 20921-400, Brazil\\
$^{16}$ Department of Astronomy, University of Illinois at Urbana-Champaign, 1002 W. Green Street, Urbana, IL 61801, USA\\
$^{17}$ National Center for Supercomputing Applications, 1205 West Clark St., Urbana, IL 61801, USA\\
$^{18}$ Institut de F\'{\i}sica d'Altes Energies (IFAE), The Barcelona Institute of Science and Technology, Campus UAB, 08193 Bellaterra (Barcelona) Spain\\
$^{19}$ Kavli Institute for Cosmological Physics, University of Chicago, Chicago, IL 60637, USA\\
$^{20}$ Institut d'Estudis Espacials de Catalunya (IEEC), 08193 Barcelona, Spain\\
$^{21}$ Institute of Space Sciences (ICE, CSIC),  Campus UAB, Carrer de Can Magrans, s/n,  08193 Barcelona, Spain\\
$^{22}$ Centro de Investigaciones Energ\'eticas, Medioambientales y Tecnol\'ogicas (CIEMAT), Madrid, Spain\\
$^{23}$ Department of Physics, IIT Hyderabad, Kandi, Telangana 502285, India\\
$^{24}$ Excellence Cluster Universe, Boltzmannstr.\ 2, 85748 Garching, Germany\\
$^{25}$ Faculty of Physics, Ludwig-Maximilians-Universit\"at, Scheinerstr. 1, 81679 Munich, Germany\\
$^{26}$ Instituto de Fisica Teorica UAM/CSIC, Universidad Autonoma de Madrid, 28049 Madrid, Spain\\
$^{27}$ Santa Cruz Institute for Particle Physics, Santa Cruz, CA 95064, USA\\
$^{28}$ Center for Cosmology and Astro-Particle Physics, The Ohio State University, Columbus, OH 43210, USA\\
$^{29}$ Department of Physics, The Ohio State University, Columbus, OH 43210, USA\\
$^{30}$ Harvard-Smithsonian Center for Astrophysics, Cambridge, MA 02138, USA\\
$^{31}$ Department of Astronomy/Steward Observatory, 933 North Cherry Avenue, Tucson, AZ 85721-0065, USA\\
$^{32}$ Australian Astronomical Observatory, North Ryde, NSW 2113, Australia\\
$^{33}$ George P. and Cynthia Woods Mitchell Institute for Fundamental Physics and Astronomy, and Department of Physics and Astronomy, Texas A\&M University, College Station, TX 77843,  USA\\
$^{34}$ Instituci\'o Catalana de Recerca i Estudis Avan\c{c}ats, E-08010 Barcelona, Spain\\
$^{35}$ Jet Propulsion Laboratory, California Institute of Technology, 4800 Oak Grove Dr., Pasadena, CA 91109, USA\\
$^{36}$ Department of Physics and Astronomy, Pevensey Building, University of Sussex, Brighton, BN1 9QH, UK\\
$^{37}$ School of Physics and Astronomy, University of Southampton,  Southampton, SO17 1BJ, UK\\
$^{38}$ Brandeis University, Physics Department, 415 South Street, Waltham MA 02453\\
$^{39}$ Instituto de F\'isica Gleb Wataghin, Universidade Estadual de Campinas, 13083-859, Campinas, SP, Brazil\\
$^{40}$ Computer Science and Mathematics Division, Oak Ridge National Laboratory, Oak Ridge, TN 37831\\
$^{41}$ Argonne National Laboratory, 9700 South Cass Avenue, Lemont, IL 60439, USA\\
$^{42}$ Cerro Tololo Inter-American Observatory, National Optical Astronomy Observatory, Casilla 603, La Serena, Chile\\

\bsp	
\label{lastpage}
\end{document}